\documentclass[conference]{IEEEtran}
\usepackage{amsmath,amssymb}
\usepackage{graphicx}
\usepackage{caption}
\usepackage{tikz}
\usepackage{booktabs}
\usepackage{epstopdf}
\usepackage{enumerate}
\usepackage{psfrag}	
\usepackage{array}
\usepackage{multirow,hhline}
\usepackage{exscale}
\usepackage{color}
\usepackage{epsfig,subfigure}
\usepackage{cite}
\usepackage{amsthm}
\usepackage{nicefrac}

\usepackage[margin=54pt]{geometry}

\newtheorem{theorem}{Theorem}
\newtheorem{lemma}{Lemma}

\definecolor{c1}{rgb}{0,0,0} 
\definecolor{c2}{rgb}{0,0,0} 
\definecolor{c3}{rgb}{0,0,0} 
\definecolor{c4}{rgb}{0,0,0}
\definecolor{c5}{rgb}{0,0,0}
\definecolor{c6}{rgb}{0,0,0} 
\definecolor{c7}{rgb}{0,0,0}
\definecolor{c8}{rgb}{0,0,0}
\definecolor{Reason}{rgb}{0,0,0}
\begin{document}

\IEEEoverridecommandlockouts
\title{\textcolor{c1}{The Generalized} Degrees of Freedom of the Interference Relay Channel \textcolor{c1}{with} Strong Interference\thanks{This work is supported in part by the German Research Foundation, Deutsche
Forschungsgemeinschaft (DFG), Germany, under grant SE 1697/3.}}
\author{
\IEEEauthorblockN{Soheyl Gherekhloo, Anas Chaaban, and Aydin Sezgin}
\IEEEauthorblockA{Chair of Communication Systems\\
RUB, Germany\\
Email: { \{soheyl.gherekhloo, anas.chaaban, aydin.sezgin\}@rub.de}}
}
\maketitle
\begin{abstract}
The interference relay channel (IRC) under strong interference is considered. A \textcolor{c8}{high-signal-to-noise ratio ($\mathsf{SNR}$)} generalized degrees of freedom (GDoF) characterization of the capacity is obtained. \textcolor{c8}{To this end, a} new GDoF upper bound is derived based on a genie-aided approach. \textcolor{c6}{The achievability of the GDoF is based on cooperative interference neutralization.}  It turns out that the relay increases the GDoF even if the relay-destination link is weak. Moreover, in contrast to the standard interference channel, the GDoF is not a monotonically increasing function of the interference strength in the strong interference regime.

\end{abstract}
\section{Introduction} 
Information theoretic results indicate that relays increase the achievable rate of a point-to-point system~\cite{CoverElgamal}.
Even wireless networks, where interference caused by concurrent transmissions is the main challenging problem, benefit from the deployment of relays which provide multiplicative gains \textcolor{c7}{in terms of achievable rates.} 
\textcolor{c4}{A multiplicative gain can be shown by comparing the generalized degrees of freedom (GDoF) of a network with and without a relay.} The GDoF is an information theoretic measure which was introduced in the \textcolor{c5}{context} of the basic interference channel by Etkin \textit{et al.} in \cite{EtkinTseWang} and is a useful approximation for \textcolor{c5}{the} capacity of a network in the high signal-to-noise ratio (SNR) regime.
\textcolor{c4}{The benefit of a relay in the IC was also shown in\cite{ChaabanSezgin_IT_IRC} by studying the GDoF of the so-called interference relay channel (IRC), an elemental network which consists of two transmitters (TX), two receivers (RX) and a relay (see Fig.~\ref{fig:SI_Model_1})}. 
The authors of~\cite{ChaabanSezgin_IT_IRC} considered the case in which the source-relay link is weaker than the interference link. \textcolor{c7}{Complementary to \cite{ChaabanSezgin_IT_IRC}, the} goal of this work is to study the impact of a relay on the GDoF when the source-relay link is stronger than the interference link under the condition that the interference itself is strong. Thus, associated with the result in \cite{ChaabanSezgin_IT_IRC}, the characterization of the GDoF for the strong interference regime is completed. \textcolor{c4}{By comparing the GDoF of the IRC} with that of the IC, \textcolor{c4}{we observe an} increase \textcolor{c4}{in} the GDoF even if the relay-destination \textcolor{c4}{link} is weak. Even more \textcolor{c7}{surprising}, the analysis shows that in the strong interference regime the GDoF can decrease as a function of the interference strength, which is a behavior not observed in the IC.
The results are \textcolor{c4}{interesting}, given the previous results \textcolor{c4}{in}~\cite{CadambeJafar_ImpactOfRelays} indicated that the degrees of freedom (DoF) of the IRC, a special and important case of the GDoF, is not increased at all by the use of relays ($\text{DoF} = 1$).

\textcolor{c7}{For the achievability, we use} \textcolor{c4}{a transmission strategy which is a combination of decode-forward \cite{CoverElgamal}, compute-forward \cite{NazerGastpar}, and a strategy named ``cooperative interference neutralization" (CN) which is a modified version of the strategy in~\cite{MohajerDiggaviFragouliTse2008}. While in the setup considered in \cite{MohajerDiggaviFragouliTse2008}, the destinations receive interference only from the relays, in our fully connected IRC, the destinations receive interference from both the relay and the undesired transmitter. Our CN strategy is designed to deal with both interferers. \textcolor{c5}{Since} our IRC is fully connected, we \textcolor{c8}{utilize} block-Markov coding~\cite{Willems}. The relay is causal, \textcolor{c5}{and therefore,} it is only able to neutralize the interference from the previously decoded blocks. This constitutes \textcolor{c5}{yet} another major difference with \cite{MohajerDiggaviFragouliTse2008}. Moreover, \cite{MohajerDiggaviFragouliTse2008} only considered the deterministic channel. In this work, we design the CN scheme for the Gaussian channel by using nested lattice codes \cite{NazerGastpar}. These codes are used in order to enable the relay to decode the sum of codewords~\cite{NazerGastpar} which is then scaled and transmitted in such a way that reduces interference at both receivers.}

\textcolor{c7}{A new upper bound on the sum capacity is derived based on \textcolor{c8}{a} genie aided \textcolor{c8}{complementing existing upper bounds from \cite{ChaabanSezgin_IT_IRC} to fully characterize the GDoF}.}

\textcolor{c5}{The rest of the paper is organized as follows. In Section~\ref{section:model_def}, we introduce the notations and the Gaussian IRC. The main result of the paper is summarized in Section~\ref{Section:Summary}. Then, in Section~\ref{section:Upper Bounds}, the new upper bound is proved. In Section~\ref{section:Achievability Scheme}, the proposed transmission scheme is motivated by considering the linear-high SNR deterministic channel model, followed by details on the relaying strategy ``CN'' and the achievability scheme for the Gaussian case. \textcolor{Reason}{In Section~\ref{Sec:Discussion}, we discuss the reason of decreasing behavior of the GDoF versus interference strength by studying the transmission scheme in details.} Finally, we conclude in Section \ref{Section:Conclusion}.}
\section{Model Definition}
\label{section:model_def}
\textcolor{c4}{Let us first define} the notations which are used in this paper. We denote a length-$n$ sequence $(x_1,\ldots,x_n)$ by $x^n$.
The functions $C(x)$ and $C^+(x)$ are defined as 
\begin{align}
C(x) = \nicefrac{1}{2}\log(1+x), \quad
C^+(x) = \left(C(x)\right)^+, 
\end{align}
where $(x)^+ = \max\lbrace0,x	\rbrace$.
A Gaussian distribution with mean $\mu$ and variance $\sigma^2$ is denoted as $\mathcal N (\mu,\sigma^2)$.
\subsection{System Model}
\label{section:system_model}
\begin{figure} 
\centering
\includegraphics[scale=0.07]{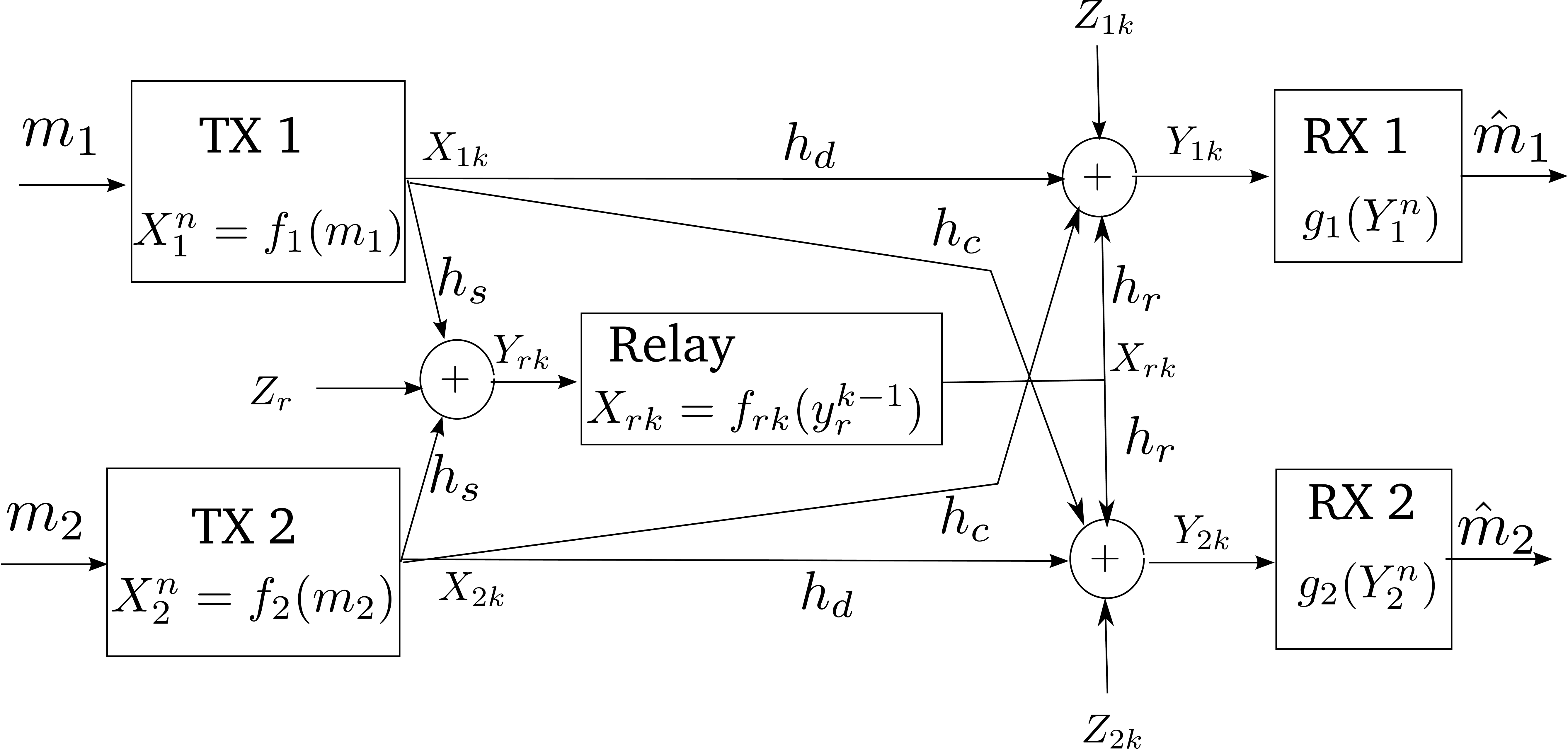}
\caption{System model for the symmetric Gaussian IRC .}
\label{fig:SI_Model_1}
\end{figure}

\textcolor{c4}{The information theoretic model of the IRC is shown in Fig.~\ref{fig:SI_Model_1}.}
Transmitter $i$ (TX$_\text{i}$), $i \in \{1,2\}$, has a message $m_i$ which is a random variable uniformly distributed over the set $\mathcal M_i \triangleq \lbrace 1,\ldots,2^{nR_i} \rbrace$ for its respective receiver (RX$_\text{i}$). The message is encoded into a codeword $x_i^n = f_i(m_i)$, where $x_{ik}$, \textcolor{c1}{$k=1,\ldots,n$}, is a realization of a real valued random variable $X_{ij}$. The transmitters must satisfy a power constraint given by 
\begin{align}
\textcolor{c7}{\frac{1}{n}\sum_{j= 1}^n\mathbb{E}[X_{ij}^2] \leq P.}
\end{align}
\textcolor{c4}{In time instant $k$, the relay receives}
\begin{align}
y_{rk} &= h_{s} x_{1k} + h_{s} x_{2k} + z_{rk} ,
\end{align}
\textcolor{c4}{where $h_s$ denotes the real valued channel gain of the source-relay channel. Moreover, $z_{rk}$ represents the additive Gaussian noise at the relay with zero mean and unit variance ($Z_r \sim \mathcal{N}(0,1)$).} The relay is causal, which means that the transmitted symbol $x_{rk}$ at time instant $k$ is a function of the received signals at the relay in the previous time instants, i.e. $x_{rk} = f_{rk}(y_r^{k-1})$. The average transmit power of the relay cannot exceed $P$. The received signals at the destinations are given by
\begin{align}
y_{jk} &= h_d  x_{jk} + h_c x_{lk} + h_{r}  x_{rk} + z_{jk}, \quad j \neq l  
\end{align}
where $j,l \in \{1,2 \}$,  and $h_d$, $h_c$, $h_{s}$, and $h_r$ represent the real valued channel gains of \textcolor{c1}{the} desired, interference, source-relay\textcolor{c1}{,} and relay-\textcolor{c1}{destination} channels\textcolor{c1}{,} respectively. The additive noise at the receivers is $Z_j \sim \mathcal N(0,1)$.
\textcolor{c2}{The probability of error, achievable rates $R_1$, $R_2$, capacity region $\mathcal C$ are defined in the standard Shannon sense~\cite{CoverThomas}.}
The sum capacity is the maximum achievable sum-rate which is given by
\begin{align}
C_{\Sigma} = \max_{(R_1,R_2)\in \mathcal C} R_\Sigma, 
\end{align}
where $R_\Sigma = R_1 + R_2$. 
\textcolor{c1}{Clearly,} the sum capacity of the channel depends on the channel gains. 

\textcolor{c4}{Since the focus of the paper is on the GDoF of the IRC, we need to define the following parameters.} \textcolor{c1}{Let $\alpha$, $\beta$, and $\gamma$ be defined as 
\begin{align}
\alpha = \frac{\log(P h_c^2)}{\log(P h_d^2)} \quad \beta  = \frac{\log(P h_r^2)}{\log(P h_d^2)} \quad \gamma = \frac{\log(P h_{s}	^2)}{\log(P h_d^2)}. \label{eq:param_def} 
\end{align}
}Then, the GDoF of the IRC, $d(\alpha, \beta , \gamma)$ is defined as
\begin{align}
d(\alpha, \beta , \gamma) = \lim_{P h_d^2\rightarrow \infty} \frac{C_{\Sigma}(\alpha,\beta,\gamma)}{\frac{1}{2}\log(P h_d^2)}\; . \label{eq:2_1}
\end{align}
\textcolor{c4}{This paper studies the IRC with strong interference $h_c^2>h_d^2$. According to \eqref{eq:param_def}, the strong interference regime corresponds to $\alpha>1$. The next section summarizes the main result of the paper.}
\section{Summary of the Main Result}
\label{Section:Summary}
In this work, we derive a new upper bound for the \textcolor{c5}{GDoF of the} IRC which is given in Lemma \ref{Lemma:new_upper_bound}. 
\begin{lemma}
\label{Lemma:new_upper_bound}
The GDoF of the IRC is upper bounded by 
\begin{align}
d \leq & \alpha + \beta. \label{eq:3_1_1}
\end{align}
\end{lemma}
The proof of the new upper bound is given in Section~\ref{section:Upper Bounds}. In addition to the new GDoF upper bound, we use some known upper bounds for the IRC which are derived in \cite{ChaabanSezgin_IT_IRC}. These upper bounds are \textcolor{c2}{restated} in Lemma \ref{Lemma:known_upper_bound}. 
\begin{lemma}
\label{Lemma:known_upper_bound}
The GDoF of the IRC is upper bounded by 
\begin{align}
d \leq  \min\lbrace& 2\max\lbrace 1,\beta \rbrace,2\max\lbrace 1,\gamma \rbrace,  \notag \\ 
&  \max\lbrace \alpha ,\beta\rbrace + (\gamma-\alpha)^+, \gamma +\alpha  \rbrace. \label{eq:known_upper_bound}
\end{align}
\end{lemma}
\begin{figure} 
\centering
\includegraphics[scale=0.4]{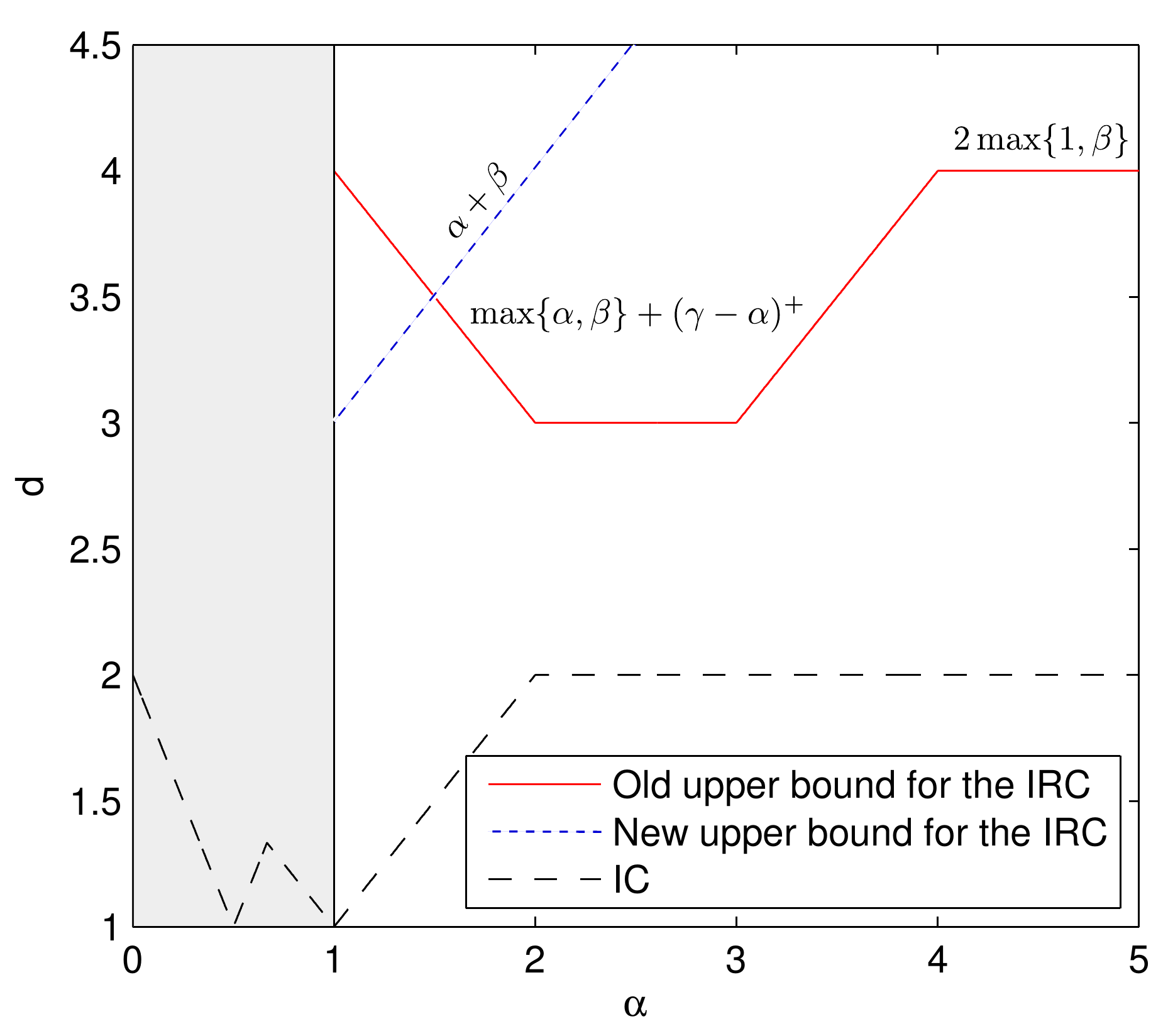}
\caption{A comparison of the GDoF of the IRC (where $\beta = 2$ and $\gamma=3$) and IC. The GDoF of the IRC is the minimum of the illustrated upper bounds. }
\label{fig:fig_comp_IRC_IC_whole_regime}
\end{figure}
Then, these upper bounds are compared with the achievable sum-rate given in Lemma \ref{Lemma:achievable_rates}\textcolor{c7}{, whose proof is deferred to Section~\ref{section:Achievability Scheme}.}
\textcolor{c2}{
\begin{lemma}
\label{Lemma:achievable_rates}
\textcolor{c2}{Let $R_{cn}^{(w)}$, $R_{cf}^{(l)}$, $R_{cm}$, and $R_{df}$ be the rates associated with the sub-messages referred to as the $w$th cooperative interference neutralization message, the $l$th compute-forward message, the common message, and the decode-forward message, respectively.}
A sum-rate $R_\Sigma$ is achievable with
\begin{align}
R_\Sigma = 2\left( \sum\limits_{w=1}^W R_{cn}^{(w)} + \sum\limits_{l=1}^L R_{cf}^{(l)} + R_{cm} + R_{df}  \right), 
\end{align}
if the constraints (\ref{eq:7_01})-\eqref{eq:7_3}, and \eqref{eq:7_12}-\eqref{eq:7_18} are satisfied under power constraints \eqref{eq:8_1} and \eqref{eq:relay_powercons}.
\end{lemma}}
\textcolor{c5}{Using the parameters in~\eqref{eq:param_def} in addition to the definition of the GDoF, we convert the sum-rate in Lemma~\ref{Lemma:achievable_rates} into the achievable GDoF of the IRC.}
Finally, by \textcolor{c5}{comparing } this achievable GDoF expression, with the upper bounds given in Lemma~\ref{Lemma:new_upper_bound} and Lemma~\ref{Lemma:known_upper_bound}, we get the GDoF in Theorem \ref{theory:GDoF_SI}. \textcolor{c5}{Notice that the GDoF of the IRC with $1\leq\alpha$ and $\gamma\leq\alpha$ is characterized completely in~\cite{ChaabanSezgin_IT_IRC}. The result for the remaining part of the strong interference regime is presented in the following Theorem.}
\begin{theorem}
The GDoF of the IRC with $1 < \alpha < \gamma$ is given by \label{theory:GDoF_SI}
\begin{align}
d = \min \{ 2 \max\lbrace 1,\beta\rbrace, \max \lbrace \alpha , \beta  \rbrace + \gamma - \alpha , \gamma + \alpha , \alpha + \beta  \} \label{eq:GDOF_SI}
\end{align}
\end{theorem}
\textcolor{c5}{In order to see the impact of the relay}, we compare the derived GDoF of the IRC with the GDoF of the IC \textcolor{c2}{in the strong interference regime given in \cite{EtkinTseWang}}
\begin{align}
d_{\mathrm{IC}}=\min\lbrace \alpha,2 \rbrace. \label{eq:9_1}
\end{align}
In \textcolor{c2}{Fig. \ref{fig:fig_comp_IRC_IC_whole_regime}, the new and the known GDoF upper bounds for the IRC and the GDoF of the IC are illustrated. As it is shown in this figure, the new upper bound is \textcolor{c7}{more binding} than the \textcolor{c5}{old} one for some values of $\alpha$. Therefore, the new upper bound is required in addition to the known upper bounds for characterizing the GDoF of the IRC. The minimum of the upper bounds gives us the GDoF of the IRC.} Moreover, \textcolor{c7}{comparing} the GDoF expression in Theorem \ref{theory:GDoF_SI} with \eqref{eq:9_1}, we conclude that the GDoF performance of the IRC is better than the IC. This increase is also obtained even if the relay-destination link is weak ($\beta<1$) \textcolor{c5}{(cf. \eqref{eq:GDOF_SI}).} The other important \textcolor{c1}{observation} is \textcolor{c1}{the} decreasing \textcolor{c1}{behavior of} the GDoF versus $\alpha$ in \textcolor{c1}{some cases}. This \textcolor{c1}{observation} is interesting because, \textcolor{c1}{to the authors' knowledge, this is the first case where a decreasing GDoF behavior is observed in the strong interference regime. This is in contrast to the IC and X-channel with strong interference where the GDoF is a nondecreasing function \textcolor{c5}{of $\alpha$} \cite{EtkinTseWang}, \cite{Chiachi_Cadambe_Jafar}}. \textcolor{Reason}{The reason of this behavior can be understood by studying the transmission scheme in the discussion in Section \ref{Sec:Discussion}.}
\section{New Upper Bound (Proof of Lemma
\ref{Lemma:new_upper_bound})}
\label{section:Upper Bounds}
In this section, we prove the upper bound given in Lemma \ref{Lemma:new_upper_bound}. To do this, we give $S^n = h_r X_r^n +Z^n$ as side information to both receivers, where $Z^n$ is \textcolor{c1}{i.i.d. $\mathcal N (0,1)$}, independent of all other random variables. Moreover, we give $Y_1^n$ and $m_1$ to receiver 2. Then, using Fano's inequality, the chain rule, \textcolor{c1}{and the independence of $m_1$ and $m_2$,} we write
\begin{align}
&n(R_1+R_2-\epsilon_n) \\ 
&\leq I(m_1;Y_1^n,S^n) + I(m_2;Y_2^n,S^n,Y_1^n,m_1) \\
& = I(m_1;S^n) + I(m_1;Y_1^n|S^n)+ I(m_2;m_1) \notag\\ &\quad + I(m_2;S^n|m_1)  
+ I(m_2;Y_1^n|S^n,m_1) \notag\\ & \quad+ I(m_2;Y_2^n|S^n,m_1,Y_1^n) \\ 
& = I(m_1,m_2;S^n) + I(m_1,m_2;Y_1^n|S^n)  \\ 
&\quad+ I(m_2;Y_2^n|S^n,m_1,Y_1^n). \label{eq:App_B_1}
\end{align}
Now, consider every term in (\ref{eq:App_B_1}) separately. The first term in (\ref{eq:App_B_1}) can be rewritten as
\begin{align}
I(m_1,m_2;S^n) & \leq I(m_1,m_2, X_r^n;S^n) \\
& \textcolor{c1}{\leq} n C(Ph_r^2). \label{eq:App_B_4}
\end{align}
The second term in (\ref{eq:App_B_1}) is given by 
\begin{align}
&I(m_2,m_1;Y_1^n|S^n)  \\ 
&\leq I(m_2,m_1,X_r^n;Y_1^n|S^n)  \\
& = h(Y_1^n|S^n) - h(Y_1^n|S^n,m_1,m_2,X_r^n)  \\
& = h(Y_1^n-S^n|S^n) - h(Y_1^n-h_rX_r^n|S^n,m_1,m_2,X_r^n)  \\
& = h(h_d X_1^n+h_c X_2^n+Z_1^n-Z^n|h_rX_r^n+Z^n) \\
& \quad- h(Z_1^n|h_rX_r^n+Z^n,m_1,m_2,X_r^n)  \\
&\overset{(a)}{\leq} h(h_d X_1^n+h_c X_2^n+Z_1^n-Z^n) - h(Z_1^n)  \\
&\leq n C\left(\textcolor{c1}{1+}P\left(h_d^2+h_c^2\right)\right),\label{eq:App_B_3}
\end{align}
\textcolor{c3}{where in ($a$), we dropped the \textcolor{c7}{conditioning} in \textcolor{c7}{the} first term because \textcolor{c7}{it does not increase} the entropy. Moreover, in the second term in ($a$), we dropped the conditions because they are all independent from $Z_1^n$.} Finally, the third term is rewritten as 
\begin{align}
& I(m_2;Y_2^n|S^n,m_1,Y_1^n) \\ 
&=h(Y_2^n|S^n,m_1,Y_1^n) -h(Y_2^n|S^n,m_1,Y_1^n,m_2) \\
& = h(Y_2^n-S^n| S^n,m_1,Y_1^n-S^n ) 
- h(Y_2^n| S^n,m_1,Y_1^n,m_2 ) \notag \\
&\overset{(a)}{\leq} h(h_dX_2^n + Z_2^n - Z^n| h_c X_2^n + Z_1^n -Z^n) - h(Z_2^n )  \\
& \overset{(b)}{\leq} h\left(h_dX_2^n + Z_2^n - Z^n- \frac{h_d}{h_c}\left(h_c X_2^n + Z_1^n -Z^n\right)\right)\notag  \\&\quad - h(Z_2^n) \notag  \\
& \leq h \left(Z_2^n-\frac{h_d}{h_c}Z_1^n+\left(\frac{h_d}{h_c}-1\right)Z^n\right) - h(Z_2^n)  \\
& \leq nC\left(\frac{h_d^2}{h_c^2} + \left(\frac{h_d}{h_c}-1\right)^2\right). \label{eq:App_B_2}
\end{align}
Since conditioning \textcolor{c7}{does not increase} entropy, we drop some conditions in the first term of $(a)$ and $(b)$. Moreover, we remove the conditions in the second term of $(a)$ because they are independent from $Z_2^n$.
Substituting the results in (\ref{eq:App_B_4}), (\ref{eq:App_B_3}), and (\ref{eq:App_B_2}) into (\ref{eq:App_B_1}), we obtain
\begin{align}
R_1 + R_2 & \leq C\left(\frac{h_d^2}{h_c^2} + \left(\frac{h_d}{h_c}-1\right)^2\right) \notag \\
& \quad+ C\left(1+ P \left(h_d^2+h_c^2\right)\right)  + C(Ph_r^2) \label{eq:new_sumrate}.
\end{align}
\textcolor{c2}{Then, using the definition of the GDoF and the parameters $\alpha$, $\beta$, and $\gamma$} in~\eqref{eq:new_sumrate} results in~\eqref{eq:3_1_1}, which concludes the proof.
\section{Achievability Scheme (Proof of Lemma \ref{Lemma:achievable_rates})}
\label{section:Achievability Scheme}
In order to show Lemma \ref{Lemma:achievable_rates}, we \textcolor{c7}{use} cooperative interference neutralization (CN). Before, explaining the CN strategy, we summarize the transmission scheme in the following deterministic example.
\subsection{A Toy Example:}
\begin{figure*}
\centering
\includegraphics[scale=0.5]{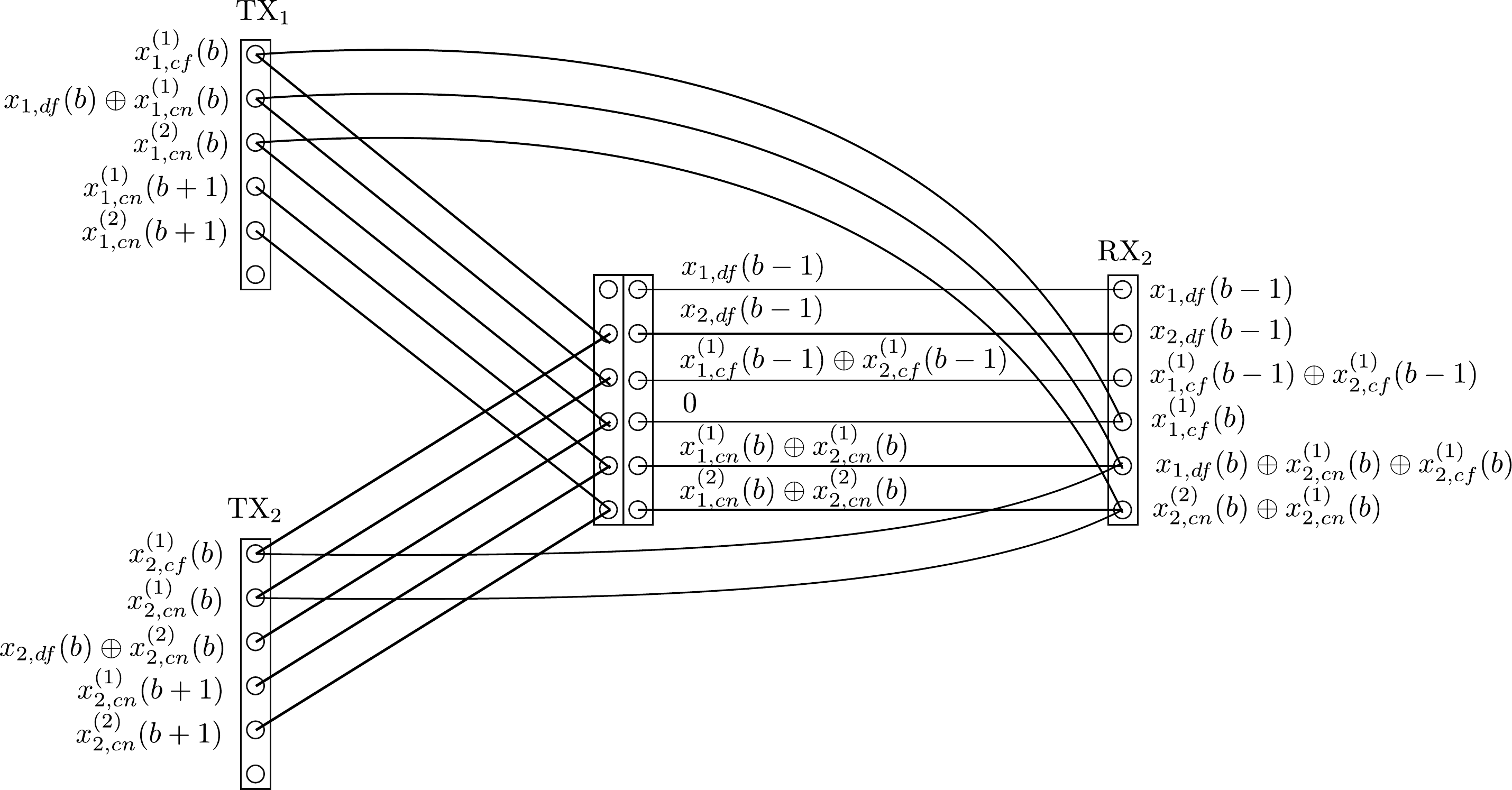}
\caption{An example for linear deterministic IRC with $n_d = 2$, $n_c = 3$, $n_r=6$, and $n_s=5$. The scheme is shown for time slot $b$. Only RX$_2$ is shown \textcolor{c5}{for} clarity.}
\label{fig:example}
\end{figure*}
\textcolor{c3}{For the sake of simplicity, we present an example based on a linear-deterministic (LD)~\textcolor{c7}{\cite{AvestimehrDiggaviTse}} IRC. The input-output relations of the LD-IRC are
\begin{align}
\boldsymbol{y}_j & = \boldsymbol{S}^{q-n_d} \boldsymbol{x}_j\oplus \boldsymbol{S}^{q-n_c} \boldsymbol{x}_l \oplus \boldsymbol{S}^{q-n_r} \boldsymbol{x}_r, \quad j\not= l  \\
\boldsymbol{y}_r & = \boldsymbol{S}^{q-n_s} (\boldsymbol{x}_1\oplus \boldsymbol{x}_2), 
\end{align}
where $\boldsymbol{x}_i$ and $\boldsymbol{y}_j$ are binary input and output vectors of length $q =\max\{n_d,n_c,n_r,n_s\}$. Here, $\boldsymbol{S}$ is a $q\times q$ shift matrix and $n_d$, $n_c$, $n_r$, and $n_s$ represent the desired, interference, relay-destination, and source-relay channels, respectively. For more information about the LD model, the reader is referred to~\cite{AvestimehrDiggaviTse}.}

In this example (Fig.~\ref{fig:example}), we fix $n_d = 2$, $n_c = 3$, $n_r=6$, and $n_s=5$. All transmitted and received vectors in time slot $b$ are given in~Fig. \ref{fig:example}. \textcolor{c5}{The transmit vector of TX$_1$ includes the information of}
\begin{itemize}
\item one CF bit
\item two current CN bits denoted by time index $(b)$
\item DF bit
\item two future CN bits represented by their time index e.g. $(b+1)$.
\end{itemize}

Since the sum of the future CN bits ($b+1$) are received at the two lower-most bits at the relay, the sum of current CN bits ($b$) is always known at the relay from the previous time slot ($b-1$). Therefore, at time slot $b$ the relay knows $x_{1,cn}^{(w)}(b)\oplus x_{2,cn}^{(w)}(b) $, where $w\in\{1,2\}$. Using this sum, we can remove the contribution of $x_{1,cn}^{(w)}(b)$ and $x_{2,cn}^{(w)}(b)$ from $y_r^n(b)$. Therefore, the relay can decode
\begin{itemize}
\item the sum of the CF bits: $x_{1,cf}^{(1)}(b)\oplus x_{2,cf}^{(1)}(b) $ 
\item the sum of future CN bits: $x_{1,cn}^{(w)}(b+1)\oplus x_{2,cn}^{(w)}(b+1) $, $w\in\{1,2\}$ 
\item the DF bits: $x_{1,df}(b)$, $x_{2,df}(b)$.
\end{itemize}
The relay forwards these known bits in the next time slot in the order shown in Fig.~\ref{fig:example}.

The receivers use backward decoding. Assuming that the decoding process of $\boldsymbol{y}_{2}(b+1)$ is successful at RX$_2$, the receiver is able to obtain
\begin{itemize}
\item $x_{1,df}(b)$ 
\item $x_{1,cf}^{(1)}(b)\oplus x_{2,cf}^{(1)}(b) $
\end{itemize}
In the next step, RX$_2$ decodes the first three bits of $\boldsymbol{y}_2(b)$. While $x_{2,df}(b)$ is desired for RX$_2$, the other ones are required in the next decoding step for interference cancellation. The receiver decodes $x_{1,cf}^{(1)}(b)$ and adds it to $x_{1,cf}^{(1)}(b)\oplus x_{2,cf}^{(1)}(b)$ to obtain the desired bit $x_{2,cf}^{(1)}(b)$. Next, the receiver removes the interference of $x_{2,cf}^{(1)}(b)$ and $x_{1,df}(b)$ from $\boldsymbol{y}_2(b)$ and decodes $x_{2,cn}^{(1)}(b)$ which is also desired. Finally, the contribution of $x_{2,cn}^{(1)}(b)$ is removed from the last bit of vector $\boldsymbol{y}_2(b)$ and $x_{2,cn}^{(2)}(b)$ \textcolor{c5}{is decoded}. \textcolor{c7}{Due to the symmetry, RX$_1$ does the same decoding process. Notice that the receivers decode the CN bits successively bit by bit. This will lead to the idea of rate splitting of the CN message in the Gaussian case considered in the next subsection.} 

\subsection{Cooperative interference neutralization:}
\label{subsection:CN_splitting_strategy}
Cooperative interference neutralization (CN) is a relaying strategy which \textcolor{c1}{was} introduced recently in~\cite{ChaabanSezginTuninetti_Journal}, \cite{ChaabanSezginTuninetti_Butterfly_Network} and \cite{ChaabanGherekhlooSezgin_SPAWC}. In this strategy, the transmitters and the relay transmit in such a way that the interference from the undesired transmitter is neutralized at the receiver. 

We introduce rate splitting to the original CN strategy~\cite{ChaabanSezginTuninetti_Journal}. For the sake of simplicity, we discuss a CN strategy with only two splits. Consider a block of transmission $b$, where $b \in \lbrace 0,\ldots,B \rbrace$ for some $B\in \mathbb{N}$. TX$_1$ wants to send the messages $m_1(1)$, $\ldots$, $m_1(B)$ in $B \in \mathbb N$ blocks of transmission to RX$_1$.
First, TX$_{\text 1}$ splits its message $m_1(b)$ into two parts, i.e. $m_{1,cn}^{(1)}(b)$ and $m_{1,cn}^{(2)}(b)$ and then encodes them using nested lattice codes. TX$_1$ and TX$_2$ use the same nested-lattice codebook $(\Lambda_{f,cn}^{(w)},\Lambda_{c,cn}^{(w)})$ with rate $R_{cn}^{(w)}$ and power $P_{cn}^{(w)}$, \textcolor{c1}{where $\Lambda_{c,cn}^{(w)}$ denotes the coarse lattice, $\Lambda_{f,cn}$ \textcolor{c5}{denotes} the fine lattice, and $w$ is the split index ($w\in\{1,2\}$).}  \textcolor{c1}{For more details about nested lattice-codes, the reader is referred to \cite{NazerGastpar}, \cite{WilsonNarayananPfisterSprintson} and \cite{ErezZamir}.} \textcolor{c1}{The} \textcolor{c5}{transmitters} encode their messages into length-$n$ codewords $\lambda_{i,cn}^{(w)}(b)$ from the nested lattice code $(\Lambda_{f,cn}^{(w)},\Lambda_{c,cn}^{(w)})$. Then, they construct the following signals
\begin{align}
x_{i,cn}^{(w),n}(b) & = \left(\lambda_{i,cn}^{(w)}(b) - d_{i,cn}^{(w)}\right) \text{ mod } \Lambda_{c,cn}^{(w)},
\end{align}
\textcolor{c1}{where $d_{i,cn}^{(w)}$ is $n$-dimensional random
dither vector.} \textcolor{c4}{Since the length of all sequences in the paper is $n$, we drop the superscript $n$ in the rest of the paper \textcolor{c5}{since it is clear from the context}. }The transmitted signal by TX$_\text{1}$ is given by
\begin{align}
x_{1}(b) & = \sum_{w=1}^2 x_{1,cn}^{(w)}(b)+\sqrt{\frac{P_{cnF}^{(w)}}{P_{cn}^w}}  x_{1,cn}^{(w)}(b+1),
\end{align}
where $b = 1,\ldots, B-1$, and $P_{cnF}^{(w)}$ denote the power of the future signal of the $w$th split, respectively. Notice that we need to consider an initialization block ($b=0$) in which the transmitter sends only the future information. Moreover, in the last block $b=B$, the users send only their current information. The other user constructs the transmit signals in the same way. 
The relay is interested only in the modulo-sum of the future CN codewords, which is 
\begin{align}
(\lambda_1^{(w)}(b+1) + \lambda_2^{(w)}(b+1)) \text{ mod } \Lambda_{c,cn}^{(w)} \label{eq:modulo_sum}
\end{align}
in block $b$.
Let us assume that the decoding process at the relay was successful in block $b-1$. Therefore, the modulo-sum of the current codewords 
is known at the relay \textcolor{c1}{at the end of block $b-1$}. The relay constructs $h_s ( x_{1,cn}^{(w)}(b) + x_{1,cn}^{(w)}(b))$ from $(\lambda_1^{(w)}(b) + \lambda_2^{(w)}(b)) \text{ mod } \Lambda_{c,cn}^{(w)}$ as shown in~\cite{Nazer_IZS2012}. \textcolor{c5}{Then}, the relay removes it from the received signal \textcolor{c1}{in block $b$}. 
\textcolor{c5}{Next}, the relay decodes the modulo-sum of the future codewords corresponding to $w=1$ and then for $w=2$ successively as follows. First, sum of the signals corresponding to $w=1$ is decoded while treating the signals $w=2$ as noise. Then, the relay removes the interference caused by $w=1$. Next the relay decodes the sum of the signals $w=2$. Using the result of \cite{NamChungLee_IT}, we conclude that the relay can decode the sum of the future CN codewords successively, if the rate satisfies
\begin{align}
R_{cn}^{(w)} \leq C^+\left(\frac{P_{cnF}^{(w)} h_s^2}{\sum_{i=w+1}^2 2 P_{cnF}^{(i)}h_{s}^2 + 1}-\frac{1}{2} \right). \label{eq:relay_dec.sum_lattice}
\end{align}
The decoded mod-$\Lambda_{c,cn}^{(w)}$ sum has power $P_{cn}^{(w)}$ as the original nested-lattice code. In every block $b = 1,\ldots, B$, the relay sends
\begin{align}
x_{r}(b) = -\sum_{w=1}^2 \frac{h_c}{h_r} \underbrace{\left[(\lambda_{1,cn}^{(w)}(b) + \lambda_{2,cn}^{(w)}(b)) \text{ mod } \Lambda_{c,cn}^{(w)} \right]}_{-x_{r,cn}^{(w)}(b)}. \label{eq:relay_sig_CN}
\end{align}
RX$_1$ wants to decode $\lambda_1^{(w)}(b)$ 
by performing backward decoding. Assume now that the future desired CN signal is decoded successfully and is known at the destination. Thus, RX$_1$ removes it from the received signal, and then divides the remaining signal by $h_c$ and adds the dither $d_{2,cn}^{(w)}$. Then, it calculates the quantization error with respect to $\Lambda_{c,cn}^{(w)}$. Similar to the decoding at the relay, the destination decodes the codeword corresponding to the first split, and then after removing its interference, it decodes the codeword of the second split. The decoding of $\lambda_{1,cn}^{(1)}(b)$ is as follow
\begin{align}
&\left(\frac{y_1}{h_c}+d_2\right) \text{ mod } \Lambda_{c,cn}^{(1)} \\
&= \textcolor{c3}{\left[x_{2,cn}^{(1)}(b) + x_{r,cn}^{(1)}(b) +  \tilde{y}_{1,cn}^{(1)}(b) + d_{2,cn}^{(1)}\right] \text{ mod } \Lambda_{c,cn}^{(1)}} \notag \\
&=\left[\left( \lambda_{2,cn}^{(1)}(b)-d_{2,cn}^{(1)} \right) \text{ mod } \Lambda_{c,cn}^{(1)} \right.  \\ 
&\quad\left. - \left(\lambda_{1,cn}^{(1)}(b) + \lambda_{2,cn}^{(1)}(b)\right) \text{ mod } \Lambda_{c,cn}^{(1)} \right.   \\
 & \quad \left.+ \tilde{y}_{1,cn}^{(1)}(b) +d_{2,cn}^{(1)}  \right] \text{ mod }\Lambda_{c,cn}^{(1)}  \\ 
&=\left[-\lambda_{1,cn}^{(1)} + \tilde{y}_{1,cn}^{(1)}(b) \right] \text{ mod }\Lambda_{c,cn}^{(1)}. \label{eq:lamba_1_decode}
\end{align}
where \textcolor{c3}{$\tilde{y}_{1,cn}^{(1)}(b)$ is the remaining part of the received signal given in}~\eqref{eq:tildenois} \textcolor{c5}{at} the top of the next page. 
\begin{figure*}
\begin{align}
\tilde{y}_{1,cn}^{(1)}(b) & = \sum_{w=1}^2\frac{h_d }{h_c }x_{1,cn}^{(w)}(b) + \left[x_{2,cn}^{(2)}(b)+ x_{r,cn}^{(2)}(b)\right] + \sum_{w=1}^2 \sqrt{\frac{P_{cnF}^{(w)}} {P_{cn}^{(w)}}}x_{2,cn}^{(w)}(b+1)    + \frac{1}{h_c}z_1(b). \label{eq:tildenois}
\end{align}
\hrule
\end{figure*}
In this way, RX$_1$ can decode $\lambda_{1,cn}^{(1)}(b)$ successfully if the rate constraint in \eqref{eq:rate_constr_rx_split} is satisfied with $w=1$.
\begin{align}
&R_{cn}^{(w)} \label{eq:rate_constr_rx_split}\leq \\ & C\left(\frac{ P_{cn}^{(w)}h_c^2}{\sum\limits_{i=w}^2 P_{cn}^{(i)}h_d^2 + h_c^2 \left(2\sum\limits_{i=w+1}^2 P_{cn}^{(i)} +  \sum\limits_{i=1}^2 P_{cnF}^{(w) } \right) +1}-1\right) \notag
\end{align}	
After decoding $\lambda_{1,cn}^{(1)}(b)$, the signal $\tilde{y}_{1,cn}^{(1)}(b)$ can be reconstructed as follows 
\begin{align}
&\left[ [-\lambda_{1,cn}^{(1)}(b) + \tilde{y}_{1,cn}^{(1)}(b)] \text{ mod } \Lambda_{c,cn}^{(1)} + \lambda_{1,cn}^{(1)} \right] \text{ mod } \Lambda_{c,cn}^{(1)} \notag \\
&=\left[  \tilde{y}_{1,cn}^{(1)}(b)\right] \text{ mod } \Lambda_{c,cn}^{(1)} = \tilde{y}_{1,cn}^{(1)}(b), 
\end{align}
where the last equality holds with high probability for some power allocations $P_{cn}^{(1)}\geq P_{cn}^{(2)}$ \cite{PhilosofZamirErezKhisti}. By using $\tilde{y}_{1,cn}^{(1)}(b)$, RX$_1$ decodes the second CN split with the rate constraint in \eqref{eq:rate_constr_rx_split} where $w=2$. Then, RX$_\text 1$ proceeds backwards till block $1$. 
\subsection{Overall transmission scheme:}
The overall transmission scheme is a combination of \textcolor{c1}{CN, CF, and DF} with the appropriate power allocation. Consider a block of transmission $b$, where $b\in\lbrace 0,\ldots,B \rbrace$ for some $B\in\mathbb N$.
\subsection{Message splitting:}
First, TX$_{\text 1}$ splits its message $m_1(b)$ as follows:
\begin{itemize}
\item a decode-forward message $m_{1,df}(b)$ \textcolor{c1}{ with rate $R_{df}$}, which is treated as in \cite{MaricDaboraGoldsmith_IT};
\item a common message $m_{1,cm}(b)$ \textcolor{c1}{ with rate $R_{cm}$}, which is treated as in a multiple access channel at the destinations;
\item $W$ CN messages $m_{1,cn}^{(w)}(b)$ \textcolor{c1}{with rate $R_{cn}^{(w)}$}, where $w=1,\ldots W$; 
\item $L$ compute-forward messages $m_{1,cf}^{(l)}(b)$; \textcolor{c1}{with rates $R_{cf}^{(l)}$}, where $l=1,\ldots,L$. These messages are treated as in \cite{ChaabanSezgin_IT_IRC}.
\end{itemize}
\subsection{Encoding}
The DF message $m_{1,df}$ is encoded using a Gaussian random code with a power $P_{df}$ into $x_{1,df}$. 
Similarly, the common message $m_{1,cm}$ is encoded using a Gaussian random code with a power $P_{cm}$ into $x_{1,cm}$.
Each CN message $m_{1,cn}^{(w)}$ is encoded into $x_{1,cn}^{(w)}$ using a nested-lattice code ($\Lambda_{f,cn}^{(w)},\Lambda_{c,cn}^{(w)}$) \textcolor{c1}{with} power $P_{cn}^{(w)}$. Moreover, each CF message $m_{1,cf}^{(l)}$ is encoded into $x_{1,cf}^{(l)}$ using a nested-lattice code ($\Lambda_{f,cf}^{(l)},\Lambda_{c,cf}^{(l)}$) with power $P_{cf}^{(l)}$. TX$_{2}$ performs the same encoding using the same nested-lattice codebooks. The signal sent by TX$_{\text 1}$ in block $b\in\lbrace 1,\ldots, B-1 \rbrace$ is given by
\begin{align}
x_{1}(b) = &  \sum_{w=1}^{W} \left(x_{1,cn}^{(w)}(b) + \sqrt{\frac{P_{cnF}^{(w)}}{P_{cn}^{(w)}}}x_{1,cnF}^{(w)}(b+1) \right)  \\  & +x_{1,df}(b)+ x_{1,cm}(b)  +\sum_{l=1}^{L}x_{1,cf}^{(l)}(b), 
\end{align}
\textcolor{c1}{The power constraint is satisfied if }
\begin{align}
\underbrace{\sum_{w=1}^{W} P_{cn}^{(w)}}_{P_{cn}} +\underbrace{\sum_{w=1}^{W}P_{cnF}^{(w)}}_{P_{cnF}} + \underbrace{\sum_{l=1}^{L} P_{cf}^{(l)}}_{P_{cf}}+  P_{df} +P_{cm}   \leq P. \label{eq:8_1}
\end{align}
\subsection{Relay processing} 
The relay starts by removing the contribution of the current CN signals as described in subsection~\ref{subsection:CN_splitting_strategy}.
\begin{figure*}
\begin{align}
R_{cm} &\leq C\left(\frac{ h_d^2 P_{cm}}{(h_d^2 + h_c^2)\left[P_{cf} + P_{cn}\right] + h_c^2  P_{cnF}  + h_r^2 \left[ P_{r,cf} + P_{r,cn}+ P_{r,df}  \right]+1}\right)  \label{eq:7_12} \\
2R_{cm} &\leq C\left( \frac{( h_d^2 + h_c^2 ) P_{cm}}{(h_d^2 + h_c^2)\left[P_{cf} + P_{cn}\right] + h_c^2  P_{cnF}  + h_r^2 \left[ P_{r,cf} + P_{r,cn}+ P_{r,df}  \right]+1}\right)  \label{eq:7_13}\\
2 R_{df} &\leq C\left(\frac{ h_r^2 P_{r,df}}{(h_d^2 + h_c^2)\left[ P_{cf}+ P_{cn} \right] + h_c^2  P_{cnF}  + h_r^2 \left[ P_{r,cf} +P_{r,cn}  \right]+1}\right)  \label{eq:7_14} \\
R_{cf}^{(1)} &\leq C\left(\frac{ h_c^2 P_{cf}^{(1)}}{h_d^2 \left[ P_{cf}+ P_{cn}\right] + h_c^2  \left(P_{cnF}+ P_{cn} + \sum_{l=2}^L P_{cf}^{(l)}\right)  + h_r^2 \left[ P_{r,cf} +P_{r,cn}  \right]+1}\right)  \label{eq:7_15}\\
R_{r,cf} &\leq C\left(\frac{ h_r^2 P_{r,cf}} {(h_d^2 + h_c^2)\left[\sum_{l=2}^L P_{cf}^{(l)}+ P_{cn}\right] + h_c^2  P_{cnF} + h_r^2 P_{r,cn} +1}  \right)  \label{eq:7_16}\\
R_{cf}^{(l)} &\leq C\left(\frac{h_c^2 P_{cf}^{(l)}}{(h_d^2 + h_c^2) P_{cn} + h_c^2 \left( P_{cnF} +  \sum_{i=l+1}^L P_{cf}^{(i)} \right) + h_d^2 \sum_{i=l}^L P_{cf}^{(i)} + h_r^2 P_{r,cn}+1} \right) \label{eq:7_17} \\
R_{cn}^{(w)} &\leq C\left(\frac{h_r^2 P_{r,cn}^{(w)}}{h_d^2 \sum_{i=w}^W P_{cn}^{(w)} + h_c^2 \left( P_{cnF}+ \sum_{i=w+1}^W P_{cn}^{(w)}\right)   + h_c^2  \sum_{i=w+1}^{W}P_{cn}^{(w)}+1}-1 \right) \label{eq:7_18}
\end{align}
\hrule
\end{figure*}
The relay decodes the messages in the following order $[m_{1,cm}, m_{2,cm}], u_{cf}^{(1)}, \ldots, u_{cf}^{(L)}$, $[m_{1,df},m_{2,df}], u_{cn}^{(1)}, \ldots, u_{cn}^{(W)}$, where $u_{cf}^{(j)}$, and $u_{cn}^{(j)}$ denote the modulo-sum of the CF and CN codewords corresponding to $j$th split, respectively. The rate constraints for successful decoding at the relay are given by
\begin{align}
R_{cm} & \leq C\left(\frac{h_{s}^2 P_{cm}} {2 h_{s}^2 ( P_{cf}+P_{cnF} + P_{df}) +1 }\right) \label{eq:7_01} \\ 
2R_{cm} & \leq C\left(\frac{2h_{s}^2 P_{cm}} {2 h_{s}^2 (P_{cf}+P_{cnF} + P_{df})+1}\right), \label{eq:7_02} 
\end{align} 
\begin{align}
R_{cf}^{(l)} &\leq C^+\left(\frac{h_{s}^2 P_{cf}^{(l)}}{2 h_{s}^2 (\sum_{i=l+1}^{L} P_{cf}^{(i)}+P_{cnF} + P_{df})+1} -\frac{1}{2}\right) \label{eq:7_1}\\
R_{df}  &\leq C\left(\frac{h_{s}^2 P_{df}}{2 h_{s}^2 P_{cnF}+1}\right), \quad
2R_{df} \leq C\left(\frac{2h_{s}^2 P_{df}}{2 h_{s}^2 P_{cnF}+1}\right) \label{eq:7_5} \\
R_{cn}^{(w)} & \leq C^+\left(\frac{h_{s}^2 P_{cnF}^{(w)}}{2 h_{s}^2\sum_{i=w+1}^{W}  P_{cnF}^{(w)}+1} -\frac{1}{2}\right). \label{eq:7_3}
\end{align}
\textcolor{c4}{The relay encodes the DF messages and all modulo-sum of the CF into length-$n$ codewords $x_{r,df}$ and $x_{r,cf}$ using a Gaussian random codebook with powers $P_{r,df}$, $P_{r,cf}$ and rates $2R_{df}$, $R_{r,cf}$, respectively. Moreover, $x_{r,cn}^{(w)}$ is constructed as in \eqref{eq:relay_sig_CN}.} Due to the causality, the relay sends the DF, CF, and CN signals in the next transmission block as follows
\begin{align}
x_{r}(b) &=  x_{r,cf}(b) + x_{r,df}(b) - \frac{h_c}{h_r}\sum_{w=1}^W x_{r,cn}^{(w)}(b),  
\end{align}
where $b=1, \ldots, B-1$. Moreover, the relay needs to satisfy the following power constraint 
\begin{align}
P_{r,cf} + P_{r,df} + \underbrace{\frac{h_c^2}{h_r^2}\sum\limits_{w=1}^W P_{cn}^{(w)}}_{P_{r,cn}} \leq P. \label{eq:relay_powercons}
\end{align}
\subsection{Decoding}
First, RX$_1$ starts decoding at the end of the last block $B$. 
It decodes the messages in the \textcolor{c1}{following}  order
\begin{align}
&\left[m_{1,cm}, m_{2,cm} \right]\rightarrow m_{r,df} \rightarrow  m_{2,cf}^{(1)} \rightarrow   m_{r,cf}\rightarrow  m_{2,cf}^{(2)} \ldots \notag \\ &\rightarrow m_{2,cf}^{(L)} \rightarrow m_{1,cn}^{(1)} \ldots \rightarrow m_{1,cn}^{(W)}.
\end{align}
\textcolor{c4}{Notice that, if $h_c>h_r$, RX$_1$ receives the CF signal from TX$_2$ on a higher power level than $x_{r,cf}$. Therefore, RX$_1$ needs to decode the CF message of TX$_2$ i.e. $m_{2,cf}^{(1)}$ before that of the relay $m_{r,cf}$. In the opposite case, if $h_c<h_d$, the optimal decoding order is vice versa. Therefore, the second to $L$th split of CF messages are all decoded after $m_{r,cf}$. Similar to CN, we need $L-1$ splits for CF messages to perform the successive decoding.}
The rate constraints for successive decoding at the destination are given in \eqref{eq:7_12}-\eqref{eq:7_18}.
\section{Discussion} \label{Sec:Discussion}
In this section, we highlight the reason of the decrease of the GDoF versus interference strength in some cases (see Fig.~\ref{fig:fig_comp_IRC_IC_whole_regime}). To this end, we study the optimal transmission schemes for different interference strength with $1<\alpha<\beta$ and with $\beta<\gamma$ and $\beta<2\alpha$.

\begin{figure*}
\centering
\subfigure[$2\alpha<\gamma$ and the achievable GDoF is $\beta+\alpha$.]{\includegraphics{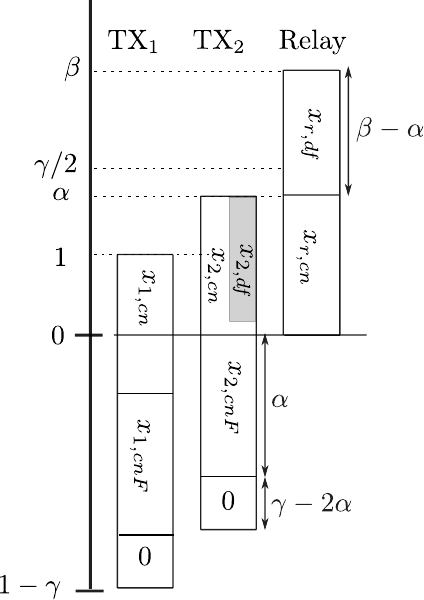} \label{fig:decreasinng_GDoF_A}} $\quad \quad \quad$
\subfigure[$2\alpha=\gamma$ and GDoF is $\alpha+\beta = \gamma+\beta- \alpha$.]{\includegraphics{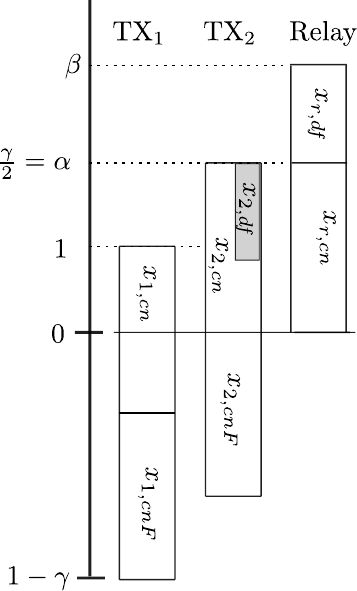} \label{fig:decreasinng_GDoF_B}}$\quad \quad \quad$
\subfigure[$2\alpha>\gamma$ and the GDoF is $\beta+\gamma-\alpha$]{\includegraphics{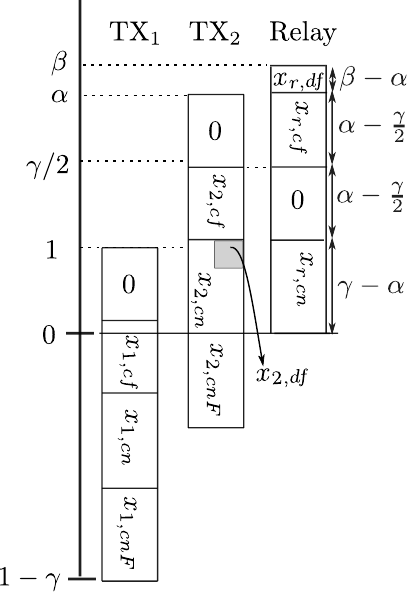} \label{fig:decreasinng_GDoF_C}}
\caption{The received signal at RX$_1$ is illustrated for three different cases when $\beta<\gamma$. The interference gets stronger from the case (a) to (c). While in (a), the transmission scheme uses the  interference to enhance the GDoF, in (c), the scheme cannot derive benefit from the increase of the interference to enhance the GDoF.}
\label{fig:decreasinng_GDoF}
\end{figure*}

First, consider the case that the capacity of the TX-relay channel is higher than twice that of the capacity of the interference channel ($\alpha<\gamma/2$). In this case, the transmission scheme is a combination of the CN and the DF strategies. From the transmission scheme, we know that the sum of current CN signals is available at the relay. Therefore, the relay is able to remove this sum before decoding the DF codeword. The relay encodes the DF codeword into $x_{r,df}$ and the sum of the CN codewords into $x_{r,cn}$. The received signal at RX$_1$ which is a superposition of the signals from TX$_1$, TX$_2$, and the relay, is shown in Fig.~\ref{fig:decreasinng_GDoF_A}. Note that the illustrations in Fig.~\ref{fig:decreasinng_GDoF} can be understood in a similar manner as in the linear deterministic model. A detailed description of such signal illustrations can be found in \cite{ChaabanSezginTuninetti_Journal}. Since $x_{r,df}$ is received at the destination on a higher power level than the interference signal, it is decoded first. 
By using backward decoding, the RX reconstructs $x_{2,df}$ from $x_{r,df}$ and cancels its interference. As it can be seen in Fig.~\ref{fig:decreasinng_GDoF_A}, the GDoF assigned to the DF signal cannot exceed $\beta-\alpha$. Moreover, it is shown in Fig.~\ref{fig:decreasinng_GDoF_A} that the relay CN signal ($x_{r,cn}$) is received on the same power level as the undesired CN signal ($x_{2,cn}$). Therefore, $x_{2,cn}$ is neutralized by the superposition with $x_{r,cn}$ and RX is able to decode its desired CN signal completely. Since in the CN strategy, we neutralize the interference signal, the GDoF of the CN signal cannot be higher than $\alpha$ (See Fig.~\ref{fig:decreasinng_GDoF_A}).

As it is shown, the relay uses its resources for neutralizing the interference (CN) and sending extra signals (DF). Roughly speaking, while a strong relay-RX channel ($\beta$) is required for forwarding extra signals, a strong TX-relay channel ($\gamma$) is needed to provide the future signals to the relay. In this region ($\alpha<\gamma/2$), the capacity of the TX-relay channel is high enough for sending all current and future signals to the relay, which can then perform as a cognitive relay. Now, suppose that the strength of the interference channel increases. Then, the TX's will use their strong channel to relay to provide more future signal (by exploiting the empty power levels under $x_{1,cnF}$ and $x_{2,cnF}$ in Fig.~\ref{fig:decreasinng_GDoF_A}). Therefore, the relay becomes more capable to neutralize the interference. While the relay will assign more power levels to neutralize the interference, the remaining power levels for extra signals (DF) will be reduced. Therefore, the GDoF of the CN signal increases while that of the DF signal decreases. Since the CN signal is desired at both users while the DF signal is desired only at RX$_2$, the overall GDoF increases versus $\alpha$. 
The increase of the GDoF stops, when $\alpha=\gamma/2$. At this point, the capacity of the TX-relay channel is exactly twice that of the interference channel. This is shown in Fig.~\ref{fig:decreasinng_GDoF_B}. Now, let the interference strength increase further. Obviously, the TX's will not be able to forward more future signal to the relay. Therefore, the relay cannot neutralize the interference completely. In order to avoid reception of the future signal ($x_{2,cnF}$) over the noise level (the $0$ level in Fig.~\ref{fig:decreasinng_GDoF_C}) and to align the CN signals of the relay with that of the undesired transmitter, we decrease the GDoF of the CN signal. Note that reducing the GDoF of the CN can cause that the GDoF of the DF signal exceeds the GDoF of the CN signal. In this case, TX$_2$ needs to assign some power levels over $x_{2,cn}$ to the DF signal which is not desired at RX$_1$. To avoid this, we need to decrease the GDoF of the DF signal as it is shown in Fig.~\ref{fig:decreasinng_GDoF_C}. By reducing the GDoF of the CN and DF signals, some empty power levels appear, which are used for adding CF signals ($x_{1,cf}$, $x_{2,cf}$, and $x_{r,cf}$ in Fig.~\ref{fig:decreasinng_GDoF_C}). While the increase of the GDoF of the CN signals compensate the decrease of that of the CF signals, reducing the GDoF of the DF signal causes a decrease in the overall GDoF versus $\alpha$ when $\gamma/2<\alpha<\beta$. 

In summery, this analysis shows that the relay uses its resources to remove the interference by neutralization and cancellation. Moreover, the remaining resources are utilized for forwarding extra signals. When the interference gets stronger, the relay reduces the GDoF of the extra signals in order to be able to remove the interference completely. This explains the non-increasing behavior of the GDoF versus interference strength in this region.
\section{Conclusion}
We characterized the GDoF of the IRC in the strong interference regime. To this end, we proposed a new upper bound for the GDoF of the IRC which is required in addition to some old upper bounds. Moreover, we suggested a transmission scheme which achieves the upper bound. This scheme is a combination of compute-forward, decode-forward, and cooperative interference neutralization. The achievability scheme is shown for a toy example based on the linear-deterministic model. The new relaying strategy ``cooperative interference neutralization" is extended for the Gaussian channel by using nested lattice codes. 
\label{Section:Conclusion}
\bibliography{root}		
\end{document}